\documentclass[twocolumn,showpacs]{revtex4}
\usepackage{graphicx}
\usepackage{dcolumn}
\usepackage{amsmath}
\begin{document}

\author{Matthew L. Wallace and B\'ela Jo\'{o}s}
\email{bjoos@uottawa.ca}
\affiliation{Ottawa-Carleton Institute for Physics, University of
Ottawa Campus,\\ Ottawa, Ontario, Canada K1N 6N5\\}

\author{Michael Plischke}
\affiliation{Department of Physics, Simon Fraser University,
\\Burnaby, British Columbia, Canada V5A 1S6}

\date{\today}

\title{The rigidity transition in polymer melts with van der Waals
interactions}

\begin{abstract}
\vspace*{0.2cm} We study the onset of rigidity near the glass
transition (GT) in a short-chain polymer melt modelled by a
bead-spring model, where all beads interact with Lennard-Jones
potentials. The properties of the system are examined above and
below the GT. In order to minimize high cooling-rate effects and
computational times, equilibrium configurations are reached via
isothermal compression. We monitor quantities such as the heat
capacity $C_{P}$, the short-time diffusion constants ${\cal D}$,
the viscosity $\eta$, and the shear modulus;  the time-dependent
shear modulus $G(t)$ is compared with the shear modulus $\mu$
obtained from an externally applied instantaneous shear. We give a
detailed analysis of the effects of such shearing on the system,
both locally and globally. It is found that the polymeric glass
only displays long-time rigid behavior below a temperature
$T_{1}$, where $T_{1}<T_{G}$. Furthermore, the linear and
non-linear relaxation regimes under applied shear are discussed.
\end{abstract}
\pacs{61.25.Hq, 66.20.+d, 64.70.Pf, 65.60.+a} \maketitle
\section{Introduction}

The onset of rigidity in disordered systems has been the focus of
a number of studies over the last decades
\cite{ThorpeBook,adam96}. At the rigidity transition, a resistance
to shear emerges, characterized by a non-zero shear modulus (above
the transition), and  a diverging viscosity (below the
transition). In an important class of systems the onset of
rigidity can be understood in terms of percolation
\cite{adam96,Barsky96,Plischke98,Plischke99,Farago02,colby,zipp1,Vernon01}.
Connectivity percolation coincides with the onset of an entropic
component to rigidity \cite{Plischke98,Plischke99,Farago02}, of
the type traditionally associated with rubber \cite{Strobl}.
Mechanical rigidity requires a rigid percolating backbone. If the
interaction forces are central this means a multiply connected
backbone, and consequently rigidity would appear at a higher
concentration of bonds than connectivity percolation \cite{Feng}.
This physical picture provides a well-defined approach to
understanding the behavior of the viscosity and the shear modulus
in the neighborhood of the onset of entropic rigidity, and has
been the focus of several recent studies
\cite{Plischke98,Plischke99,Farago02,colby,zipp1,Vernon01}.  The
onset of the mechanical rigidity arising from covalent bonds has
been extensively studied in network glasses at temperatures well
below the glass transition (GT). In a mean-field argument, this
component of rigidity sets in when the density of floppy modes
approaches zero, or the number of constraints exceeds the number
of degrees of freedom.  For bond-bending networks, this means a
mean coordination number $\langle r \rangle = 2.4$
\cite{Phillips,Thorpe83,Thorpe,Huerta02,AngellBook}.

In this paper we consider the onset of rigidity in a polymer melt
which occurs as the temperature is lowered below the GT. There is
no obvious length scale emerging a priori as the chains bind
together under the attractive van der Waals interactions between
monomers forming disordered structures. As we shall see, the
region around the glass transition, where the viscosity is
expected to diverge and a finite shear modulus to emerge, becomes
more complex than in percolation driven systems. Where precisely
the onset of rigidity occurs is an open question \cite{Huerta02}.
There is a body of literature on the behavior of the viscosity
near the glass transition which will be used to discuss our
results. There is however much less information about the onset of
rigidity as measured by the emergence of a shear modulus.

The article proceeds as follows. Section II provides some general,
conceptual background information regarding the GT, and in
particular, the behavior of the viscosity. Section III discusses
our model and the computational tools that we have used. Section
IV presents the main results obtained from these simulations and
some brief explanations. In addition to the viscosity and the
shear modulus we have calculated the diffusion constant and the
heat capacity. Section V, before the Conclusion,  discusses the
results within the current framework of polymeric glasses and
provides some insight into the theoretical concepts discussed.

\section{Rigidity and the glass transition}\label{glasstransition}

For a polymer melt,  in contrast to the rigidity that sets in with
increasing crosslink density, lowering the temperature drives the
system reversibly towards a homogeneous glass phase with no
obvious diverging length scale.  Attempts have been made to invoke
the percolation of the slow regions near the glass transition
\cite{Yilmaz02,Long01}. And also in a particular model where rigid
bonds can be formed upon cooling, a link is made to a mechanical
rigidity transition based on constraint counting
\cite{Huerta03,Narayanan99}.  We will focus on the behavior of the
viscosity and the shear modulus and analyze their critical
behavior in a simple numerical model of a polymer melt made of
short freely jointed chains. Discussions will be made using
existing models of the physical properties of the system near the
glass transition.

Historically viscosity has played an important role in analyzing
glasses. The glass transition was associated for practical
purposes with a specific number for the viscosity: $\eta =
10^{13}$ poise or $10^{12}$ Pa$\cdot$s, where the material could
be considered not to flow during experimental time scales.

The glass transition is known as a ``pseudo second-order
transition'', since the discontinuities are not sharp and occur
over a range of temperatures \cite{Rao}. As well, a glassy system
is not actually in equilibrium, because aging phenomena and flow
have a non-negligible effect at long times. At best, we refer to a
metastable equilibrium which, on our timescales (as in the case of
short experimental times), can be treated as an equilibrium
system.

A commonly-used approach to the glass transition has been
mode-coupling theory (MCT), derived from microscopic density
fluctuations within a liquid \cite{Binder03,Gotze92}. A dynamic
phase transition is predicted at approx. 1.1-1.3 times $T_{G}$,
involving a divergence in the zero shear-rate viscosity $\eta$,

\begin{equation}\label{mct}
\eta = (T - T_{MC})^{-s}.
\end{equation}
which could be masked by thermal activation effects \cite{Perez}.
More importantly, idealized MCT (which does not take into account
``hopping'' processes, for instance) predicts an
ergodicity-breaking at $T=T_{MC}$ \cite{Binder03}. It has also
been proposed that there is, in fact, some type of ``crossover
point'' to the Vogel-Fulcher-Tamman (VFT) regime (see
\cite{Binder03} and Refs. therein). The latter is a more
successful framework for analyzing glass-forming liquids and is
defined by the VFT law, an empirical equation of the form

\begin{equation}\label{etaVFT}
\eta = \eta _\infty \exp \left( \frac{E_{act} }{T - T_0 } \right).
\end{equation}
The VFT temperature $T_0$ is expected and has been observed to be
very close to the Kauzmann temperature $T_K$ discussed in Refs.
\cite{Binder03,Angell00}. $T_K$ is the temperature which defines a
hypothetical state with zero configurational entropy or the
formation of an ``ideal'' glass. Such a state is never reached
because of the large slowing-down of the system at $T_G$.
Impressive fits to the VFT equation have been achieved
\cite{Okun97,Varnik02} above the glass transition. These fits also
provide a means to characterize the fragility of glasses: the
closer $T_G$ is to $T_0$ the more fragile the glass
\cite{Angell95,Colby00}. Nevertheless, the range of applicability
of the VFT equation is under contention; some research indicates
that it cannot go as low as $T_G$, while other studies suggest
that it is best applied close to the GT (see Ref. \cite{Angell00}
for a review).

A dynamical scaling approach was recently proposed to explain the
divergence in relaxation time below the GT \cite{Colby00}. The GT
is characterized by dynamical heterogeneity, in that only some of
the material is able to move due to a lack of free volume. At a
critical temperature $T_C$, there is not enough free volume for
any particles to move and local motion is effectively prohibited.
This approach assumes percolation-like universal behavior among
glass-forming liquids with an additional activation energy term of
the following form

\begin{equation}\label{Colby}
\eta \sim \left( \frac{T - T_C}{T_C} \right)^{-9} \exp \left(
\frac{E}{kT} \right)
\end{equation}
Unlike the MCT approach, the above equation yields a critical
temperature for viscosity divergence $T_C$ below $T_G$. While
experimental data from polymeric liquids is well-described without
the activation energy contribution, the extra term ensures the
same universality class for all glass formers \cite{Colby00}.

One can also consider the GT in terms of the ``energy landscape''
of the system, a hypersurface punctuated by local minima of
varying depth, where the shape depends on the volume and the
sampling on the temperature
\cite{Rao,Angell00,Buchenau03,Debenedetti01}. In terms of energy,
$T_{G}$ can be seen as a temperature at which the system is
``trapped'' in a relatively deep local minimum and the free energy
barriers effectively prevent further exploration of phase space.
Polymeric glasses do not behave like other, more simple glasses,
such as SiO$_{2}$. Simple glasses have a more precisely defined
energy landscape, whereas polymeric glasses have more
configurational possibilities, yielding more local minima which
can be sampled via local rearrangements. As such, polymers tend to
create relatively ``fragile'' glasses, characterized by a
non-Arrhenius variation of the viscosity with temperature. In
terms of the glass transition temperature, it has been observed
that $T_{G}$ rises with chain length or molecular weight until a
plateau around $M=100$ \cite{Perez}.

In general, glasses display a more complex response to
low-frequency shearing than purely elastic systems. Nonlinear
mechanical responses are common in glass systems after
ergodicity-breaking \cite{Angell00}, and the relaxation of a
simple glass will depend on local rearrangement. In polymer
systems there is an additional component arising from the
extension of the chains. This, as well as energy landscapes, will
be useful tools in interpreting the mechanical response within our
glassy polymer melt. Some aspects of relaxation below the GT are
well understood \cite{Ediger96,Angell00} in terms of aging
subsequent to a temperature quench or in response to a mechanical
deformation, after which glasses can relax fairly quickly. However
this has not, to our knowledge, been examined using an
instantaneous simple shear experiment.

\section{Model} \label{model}

\subsection{The bead-spring model}

We have used the well-known bead-spring model based on work by
Kremer and Grest \cite{Kremer90} and more recently applied with
success by the research group of Kurt Binder \emph{et al}.
\cite{Binder03,Okun97,Bennemann99,Bennemann98,Buchholz02}. Their
work has focused on issues such as $\alpha$ and
$\beta$-relaxation, cooling-rate dependencies and cage effects on
approaching $T_G$ from above. In this model, the neighboring beads
along each chain (there are 10 monomers per chain and a total of
$N=1050$ monomers) interact through the FENE potential

\begin{equation}\label{FENE}
\mbox{U}_{FENE} (r_{ij}) = \frac{1}{2} kR_0^2 \log \left[ 1 -
\left( \frac{r_{ij}}{R_0}  \right)^2 \right]
\end{equation}

\noindent where $R_0=1.5$, $k=30$. In addition, all particles
interact though the truncated Lennard-Jones potential,

\begin{equation}\label{LJ}
U_{LJ} \left( {r_{ij} } \right) = 4\varepsilon _{LJ}\left[ {\left(
{\frac{\sigma }{r_{ij} }} \right)^{12} - \left( {\frac{\sigma
}{r_{ij} }} \right)^6} \right] + C\,\,,\,\,\,\,r_{ij} < 2.5\sigma
\,
\end{equation}

\noindent where $C$ is chosen such that the potential is zero at
the cutoff radius.

The length is rescaled using $\sigma =1$ and the temperature is
expressed in units of $\varepsilon _{LJ} /k_B$. Combining the two
potentials yields an optimum bond length of $0.96 \sigma $, which
inhibits crystallization by introducing a competing length scale.

\subsection{Approaching the glass phase along an isotherm}

Usually the GT is approached at constant $P$ by lowering the
temperature
\cite{Okun97,Bennemann99,Bennemann98,Buchholz02,Bennemann99b}. In
order to avoid problems associated with high cooling rates
\cite{Buchholz02} and to reduce computational time, we propose a
method whereby the samples can be ``frozen'' or not, as the case
may be, by compression at a desired $T$. Although we are still
measuring changes in a given quantity with respect to temperature
with a constant pressure, we are effectively using a different
thermodynamic path to define the state of a sample. There is a
very complete discussion of thermodynamic paths in Ref.
\cite{Bennemann99b}. However, it does not discuss defining a
thermodynamic state using an isothermal path. Our method seems to
allow a better exploration of phase space and consequently, a more
physically meaningful final state for low temperatures. In a pure
Lennard-Jones simulation, for instance, we found that
crystallization requires less CPU time when compressing the system
than when simply cooling it \cite{Wallace2}. The temperatures
studied range from 1.2 to 0.2 $\epsilon _{LJ} /k_B$, allowing us
to explore the liquid, supercooled and glassy regimes. We entered
the glass phase both along an isochore and an isobar.  However, we
retained only the simulation results along the isobar, as pressure
variations along the isochore proved somewhat unphysical. The main
problem was a progressive increase of pressure with lowering
temperature below $T_G$ (similar results were found in Ref.
\cite{Bennemann98}). This discrepancy effectively amounts to

\begin{equation}\label{Maxwell}
\left( {\frac{\partial P}{\partial T}} \right)_{N,V} =
\frac{T}{V}\left( {\frac{\partial S}{\partial T}} \right)_{N,V} <
0
\end{equation}

\noindent which is clearly forbidden. Along the constant volume
path, the system goes through what would have been a two-phase
region in the thermodynamic limit.

For a given sample, we create an initial ``gas'' of polymers at
the desired temperature, having ensured their separation and
random orientations. The chains reach their individual equilibrium
configurations in an NVT ensemble using an algorithm based on the
Langevin equation, called Brownian Dynamics (BD) in Ref.
\cite{Allen}:

\begin{equation}\label{Brown}
m\frac{d^2x_i }{dt^2} = -{\frac{\partial U_i}{\partial x_i}} +
m\Gamma\frac{dx_i }{dt}  - W_i(t).
\end{equation}
This algorithm allows the system to reach equilibrium in our
``expanded'' volume quickly and realistically, by simulating a
heat bath in the form of a friction coefficient $\Gamma$ and by
introducing a random force $W$ in the form of Gaussian noise.
Subsequently, the system is compressed to the pressure $P=1.0
\epsilon _{LJ} /\sigma^3$ to achieve a similar system to that of
K. Binder's research group
\cite{Okun97,Bennemann99,Bennemann98,Buchholz02,Bennemann99b}. The
compression is done in the BD NVT ensemble by reducing affinely
the system size at a specified rate. The largest compression rate
of a side of the box, in Lennard-Jones units, was around
$0.015\sqrt {\varepsilon _{LJ}/m}$. Using a smaller compression
rate was shown to cause negligible change in quantities such as
viscosity and energy. Naturally, the rate of compression will
inevitably have some effect on determining at what temperature
$T_G$ the glass becomes ``stuck.'' However, we believe that, by
continuously changing the volume (and thus the precise shape of
the energy landscape), the system can eventually find a lower
energy minimum. The final volume is established at a given
temperature for $P=1$ using a constant pressure damped-force
algorithm \cite{Allen}. The system is then allowed to evolve in a
micro-canonical ensemble by MD. We need to run in that ensemble to
collect information on the correlators required to calculate
viscosity and shear modulus related quantities. The equations of
motion were integrated with the standard velocity Verlet algorithm
\cite{Allen} and a time step of $dt=0.005\sqrt {{m\sigma ^2}
\mathord{\left/ {\vphantom {{m\sigma ^2} \varepsilon _{LJ}}}
\right. \kern-\nulldelimiterspace} \varepsilon _{LJ} } $. During
this portion of the simulation, both the pressure and the
temperature were seen to fluctuate around their expected values.

\subsection{Viscosity and shear modulus from the stress autocorrelation function \label{viscG}}

The time-dependent shear modulus $G(t)$ measures the response of
the system to a shear strain $\epsilon _{\alpha \beta}$ applied at
$t=0$, and is defined as \cite{Strobl}
\begin{equation}\label{Gdef}
G(t) = \frac{ \sigma _{\alpha \beta} (t)}{\epsilon _{\alpha
\beta}} ,\,\,\,\,\,\,\,\alpha \ne \beta.
\end{equation}
$\sigma _{\alpha \beta}$ is an off-diagonal macroscopic stress
tensor element and implicitly includes inter and intra-chain
interactions. $G(t)$ can be calculated from the stress
fluctuations in the quiescent melt using the fluctuation
dissipation theorem which states that \cite{Strobl}:

\begin{equation}\label{G}
G(t) = \frac{V}{kT} < \sigma _{\alpha \beta} (t_0 )\sigma _{\alpha
\beta} (t_0 + t)
>,\,\,\,\,\,\,\,\alpha \ne \beta
\end{equation}

\noindent where

\begin{equation}\label{stress}
\sigma _{\alpha \beta } = - \frac{1}{V} \left[ \sum\limits_{i =
1}^N m v_{i\alpha } v_{i\beta } - \sum\limits_{i < j}^N
{\frac{r_{ij\alpha } r_{ij\beta } }{r_{ij} }} \frac{\partial
U_{ij} }{\partial r_{ij} }\right].
\end{equation}
The sums are over all the particles in the system indexed from 1
to $N$. $G(t)$ is a very powerful tool: it allows us to obtain a
zero-shear limit for the shear modulus without external
deformation, which can have a non-negligible effect on the steady
state. Non-ergodic glassy materials are especially sensitive to
configurational changes. Instead, $G(t)$ simply uses the
microscopic density fluctuations and the subsequent response to
give us a clear picture of how the stress evolves in a system. At
$t \to \infty$, $G(t)$ becomes the ``equilibrium modulus,''
$G_{eq}$ and, at $t \to 0$, $G(t)$ becomes $G_{\infty}$, the
infinite-frequency modulus \cite{Lodge97,Lodge99}. To our
knowledge, $G_{eq}$ has never been compared to the usual shear
modulus, which is measured by a system's response to a shear
deformation (see Section \ref{deform} below). The zero shear-rate
viscosity is obtained by the appropriate Green-Kubo formula:

\begin{equation}\label{visc}
\eta= \int\limits_0^\infty {G(t)dt},
\end{equation}

\noindent from which one can obtain the relaxation time as

\begin{equation}\label{relax}
\tau = \frac{\eta}{G_{\infty}}.
\end{equation}

In the past, viscosity measurements have been made using
non-equilibrium molecular dynamics (NEMD) to avoid the problem of
long time tails of correlation functions at temperatures
approaching $T_{G}$  \cite{Varnik02}.  This issue was resolved by
fitting $G(t)$ to the KWW, or stretched-exponential function and
integrating to infinity, via \cite{Lodge97}

\begin{equation}\label{stretched}
G(t) = G(0)\exp \left[- {\left( {\frac{t}{t_0 }} \right)^\beta }
\right].
\end{equation}
\noindent  The quality of the fits as well as previous research
\cite{Angell00} suggests that this approach reflects well the
behavior of the system over long time scales. Although there is a
certain additional error associated with the fit, it allows us to
keep a system in its ``true'', undriven state. The errors were
large for $T$ very close to $T_G$ ($T \sim T_G + 0.02$), due to
the increasing length of the tails.

The nature of the tail changes significantly below the glass
transition, and the KWW fit of Eq. (\ref{stretched}) is no longer
appropriate. The slow relaxation of the glass through thermally
activated processes is better represented by the power law form
\cite{Lodge99,Lodge97}:
\begin{equation}\label{Gpower}
G(t) = G_{eq} + At^{ - \varphi }
\end{equation}
In physical terms the power law fit uses the distribution of
energy barriers explored by the system during the simulation time
to predict the long time behavior.

\subsection{Short-time diffusion coefficients \label{diff}}

To get an idea of the mobility of the chains and the individual
particles, we define  short time diffusion coefficients using the
Einstein relations, ${\cal D}_{M}$, for individual particles $i$,
and ${\cal D}_{CM}$, for the centers of mass of individual chains
$j$:

\begin{equation}\label{diff-mol}
{\cal D}_M = \frac{\left\langle {\left[ {{\bf r}_i \left( t
\right) - {\bf r}_i \left( 0 \right)} \right]^2} \right\rangle
}{6t}\,\,
\end{equation}

\begin{equation}\label{diff-cm}
{\cal D}_{CM} = \frac{\left\langle {\left[ {{\bf r}_j \left( t
\right) - {\bf r}_j \left( 0 \right)} \right]^2} \right\rangle
}{6t}.
\end{equation}
Ignoring initial displacements, we simply retain the slope of the
first approximately linear part of the mean-square displacement
vs. time graph, calculated around 30000$dt$. In the asymptotic
limit if the particles are delocalized, there would be on a longer
timescale a linear regime with a smaller slope. If they are
localized, $(<\Delta r>)^2$ would be bounded. In the liquid phase,
the first linear regime essentially reflects the local diffusion
of a particle, which is limited by the proximity to its neighbors.
In the glassy regime, there may be little or no diffusive
behavior. Nevertheless, the approximately constant slope is an
indicator of how much local mobility a given monomer or chain can
have within its cage.

\subsection{The shear modulus from the shear deformation \label{deform}}

We have also calculated the shear modulus directly by applying a
simple shear deformation to a given sample.  We start with a cube
of side $L$. For a given strain $\varepsilon _{xy}$, an affine
shear deformation in the $x$ direction is applied (in a plane with
its normal along the $y$ direction).   An atom initially in
position $(x,y,z)$ is displaced to $(x+ \epsilon _{xy}y, y, z)$.
The boundaries of the simulation box consequently are shifted  for
$x_{min}$ from 0 to $\varepsilon _{xy} y $ , and for $x_{max}$
from $L$ to $\varepsilon _{xy} y  + L$. This is applied as a
one-time, instantaneous deformation. The shear modulus is then
calculated with an off-diagonal element of the stress tensor, once
the system is equilibrated:

\begin{equation}\label{mu}
\mu = \frac{\left[ {\sigma _{xy} \left( \varepsilon  _{xy}\right)
- \sigma _{xy} \left( 0 \right)} \right]}{\varepsilon _{xy}}
\end{equation}

The shearing is applied in the five other directions, substituting
$xy$ by $yz$,$-xy$, etc. Individual samples lack symmetry and
therefore the various deformations will not usually give the same
stress components. The evolution of the residual stress is
monitored, until a stress plateau is reached in the deformed
system. Deformations of $\varepsilon =0.01$ to $0.2$ were
performed. A simple shear deformation is more rigorous than a pure
shear deformation, which relies heavily on the assumption of
isotropy throughout the system \cite{Plischke99}. In a pure shear
deformation, a sample would have been stretched along $x$ by
$\varepsilon _{xy}$ resulting in a compression along the two other
directions by $\varepsilon  _{xy}/2$ in a volume preserving
system.

\subsection{Heat capacity \label{Cp}}

Finally, we monitored the changes in the heat capacity $C_{P}$
across $T_{G}$. Although other methods exist for calculating
$C_{P}$ via MD simulations, the best approach was to calculate the
potential energy in the NVT ensemble (Brownian dynamics) for each
temperature. Knowing the volume and the pressure at each
temperature, we can obtain the specific heat per particle $C_P$ in
units of $k_B$ from:

\begin{equation}\label{$C_{P}$}
C_P = \frac{1}{N}\left( {\frac{\partial H }{\partial T}}
\right)_{N,P},
\end{equation}
where $H= E + PV$ is the enthalpy.

\subsection{The simulations}

Most simulations were carried out on the ``Bugaboo'' Beowulf
Cluster at Simon Fraser University, with computational times for
one complete sample at a given temperature (\textit{ab initio} and
including all correlations and shearing) at around 6 days on a
single processor. Approximately 10 samples of 1050 particles are
examined for each temperature, with each given sample being used
to calculate approximately 400 ``correlators'' of length
30000$dt$.

\section{Results}

\subsection{ The glass transition temperature $T_{G}$}

\begin{figure}
\resizebox{3.25in}{2.5in}{\includegraphics{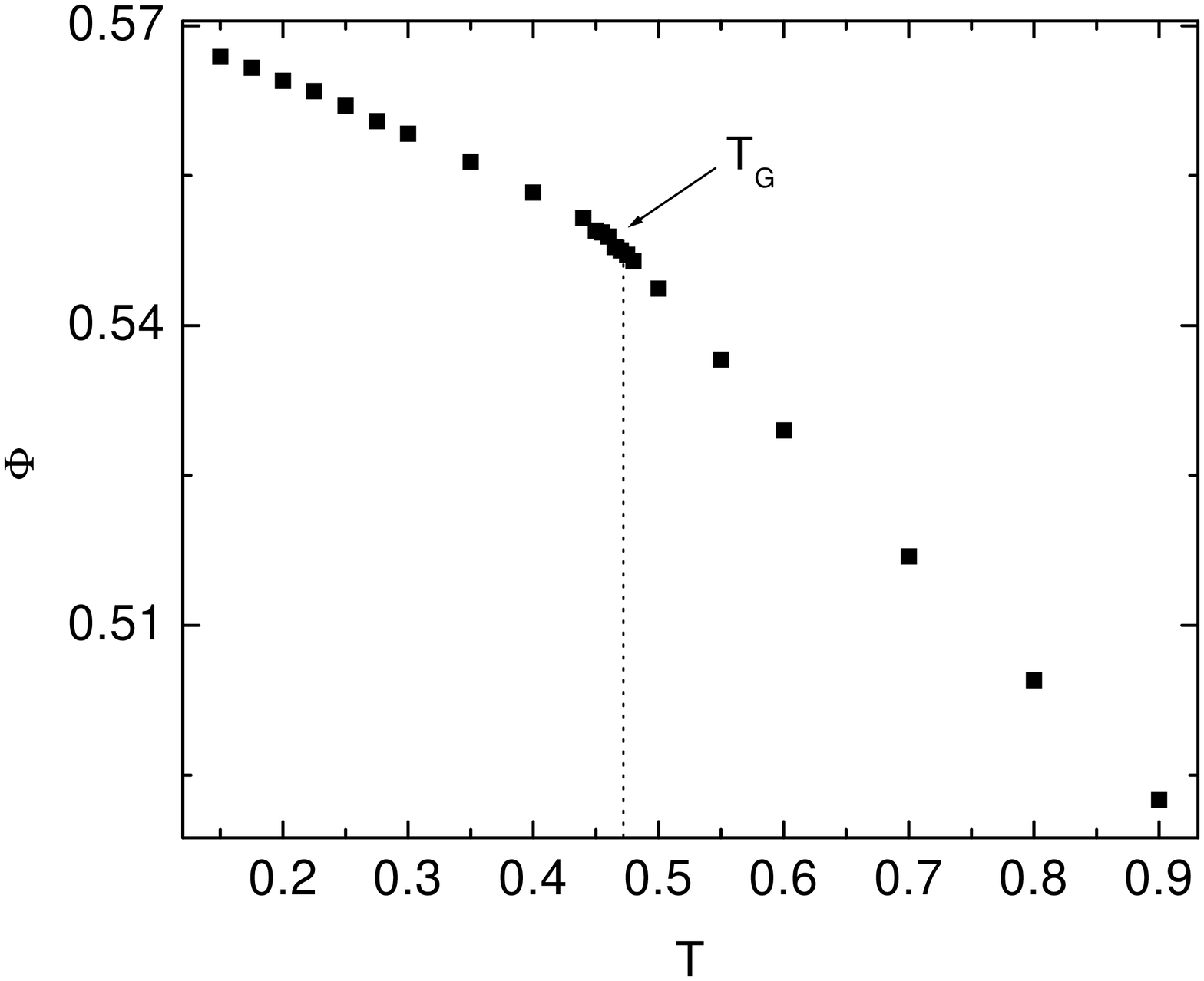}}
\caption{Packing fraction $\Phi$ as a function of temperature. The
intersection of the two slopes in the glassy and liquid regimes
accurately determines $T_{G}$, shown with the arrow and dotted
line.}\label{fig2}
\end{figure}

\begin{figure}
\resizebox{3.25in}{2.5in}{\includegraphics{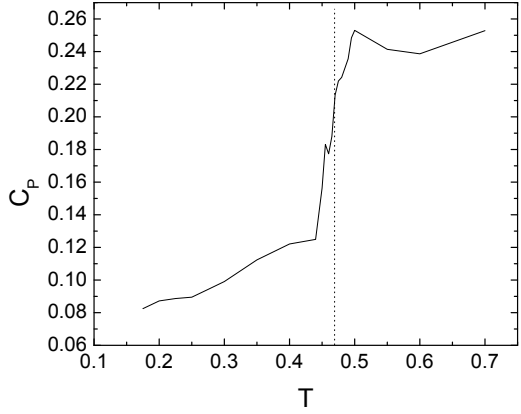}} \caption{Heat
capacity $C_{P}$ found by numerically deriving the enthalpy $H$
with respect to temperature. The dotted line shows the location of
$T_G$ as determined from Fig. \ref{fig2}. Note that the drop in
$C_P$ as T is lowered ends at $T=0.44$, coinciding with the
appearance of long-term rigidity (see Fig
\ref{fig9}).}\label{fig3}
\end{figure}

We begin by establishing the glass transition temperature $T_{G}$,
using the common and accessible method (both experimentally and by
simulation) of examining the volume (or equivalently, the packing
fraction $\Phi$) across the GT \cite{Strobl,Kruger02} along the
isobar $P=1$. The intersection of the slopes in the solid and
liquid regimes separates two regions of nearly constant volume
expansivity. It yields an unambiguous value of $T_{G}=(0.465 \pm
0.005)$, as shown in Fig. \ref{fig2} and Table I. The variation of
$C_{P}$ with respect to temperature provides another estimate. As
discussed in Section \ref{Cp}, $C_{P}$ was calculated from the
derivative of the enthalpy with respect to temperature along an
isobar (Fig. \ref{fig3}). Such an approach has been extensively
used to calculate $T_{G}$ and as expected, $C_{P}$ is
characterized by a sharp increase around $T_G$ \cite{Rao}. The
increase, however, takes place over a temperature interval, and it
is not clear what part of the curve should be used as the location
of $T_G$  (see Fig. \ref{fig3}). The start of the rapid rise
occurs at $0.44$ and ends at $0.50$. The $T_G$ determined above
from the change in value of the volume expansivity lies in the
middle of the rapid rise.  The beginning of the rapid rise
suggests that a major change in the entropy is taking place at
that point. As we shall see later, this point has a special
significance.

\subsection{The viscosity}

\begin{figure}
\resizebox{3.25in}{2.5in}{\includegraphics{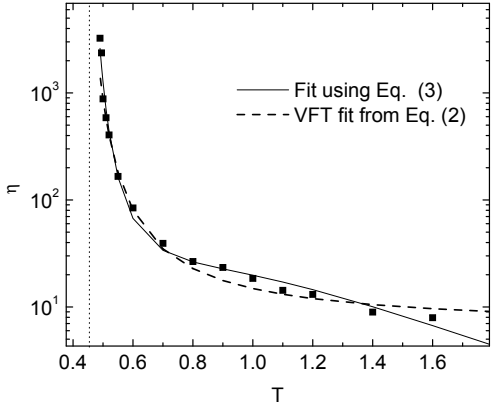}}
\caption{Viscosity $\eta$ above the glass transition, along with
the corresponding fits from Eqs. (2) and (3). The dotted vertical
line marks $T_G$ obtained previously.}\label{fig4}
\end{figure}

Now that we can identify the GT within our system, we can examine
the relevant dynamical quantities around this point. First, we
obtain the expected viscosity curve with the characteristic
increase over several orders of magnitude around $T_{G}$, along
with the corresponding curves of Eqs. (\ref{etaVFT}) and
(\ref{Colby}) obtained using a simple non-linear curve-fitting
algorithm (Fig. \ref{fig4}). We also attempt a fit using the MCT
approach in Eq. (\ref{mct}). The results are summarized in Table I
and follow the same trends to those found in Refs.
\cite{Bennemann98} and \cite{Binder03} for a similar system. The
attractive LJ interaction has a shorter cut-off $2.24 \sigma$,
instead of our $2.5 \sigma$ leading to systematically lower
critical temperatures.  Varnik and Binder \cite{Varnik02} have
looked at the increase in viscosity with lowering of the
temperature in the same system but with the melt driven by a force
field. The values of $T_{MC}$ and $T_0$ are quite different under
those conditions, expectedly lower.

\begin{table}[htbp]
\caption{Relevant temperatures characterizing the glass and
rigidity transitions: $T_G$ is the glass transition with the
associated MCT critical temperature $T_{MC}$, the VFT temperature
$T_{0}$, and the critical temperature $T_C$ obtained from Eq.
(\ref{Colby}). In addition, we present $T_{1}$, corresponding to
the onset of true rigidity, found by extrapolation to $G_{eq}$ in
Eq. (\ref{Gpower}).}
\begin{tabular}{cccccc}
\hline \hline
\\ $T_{G}$ &&& \\ 0.465 $\pm$
0.005 &&&    \\ \\  $T_{MC}$ & $T_{0}$ & $T_C$ & $T_{1}$  \\ 0.51
$\pm$ 0.02
& 0.41 $\pm$ 0.02 & 0.422 $\pm$ 0.006 & 0.44 $\pm $0.01\\
\hline
\end{tabular} \label{tab1}
\end{table}

We get an idea of fragility by plotting the log of the viscosity
$\eta$ versus the inverse temperature,  a characteristic
$T_G$-scaled Arrhenius curve for glassy systems (Fig. \ref{fig7}).
Evidence of fragility (deviations from Arrhenius behavior) becomes
apparent as we approach the GT. Typical experimental data from
polymeric glasses generally show very fragile behavior from such
plots \cite{Angell00}.

\subsection{The shear modulus from $G(t)$}

We have also examined how $G(t)$, the time-dependent shear
modulus, changes as we cross the glass transition (Fig.
\ref{fig8}). The main difference between solid and liquid regimes
occurs in the long-time tails of $G(t)$. While a KWW fit of Eq.
(\ref{stretched}) can be used for $T > T_{G}$, a power law of the
form of Eq. (\ref{Gpower}) must be applied for $T<T_{G}$
\cite{Lodge99,Lodge97}.  Simply evaluating the value of $G_{eq}' =
G(t)$ at the end of our correlation function ($t=150\sqrt
{{m\sigma ^2} \mathord{\left/ {\vphantom {{m\sigma ^2} \varepsilon
}} \right. \kern-\nulldelimiterspace} \varepsilon }$) gives a good
idea of the initial degree of relaxation possible in our system,
or the shear modulus at short timescales. However, we would like
to examine the ``true,'' long-time relaxation of the glass, which
is generally not accessible at simulation timescales. As such, we
find $G_{eq}$ (at $t\to\infty )$ using a non-linear fit to the
power-law of Eq. (\ref{Gpower}), giving non-zero values of
$G_{eq}$ beginning at $T_1=0.44 \pm 0.01$, and an approximately
linear increase with decreasing temperature, as shown in Fig.
\ref{fig9}. The decay of $G(t)$ is very slow, and most of the
information necessary for the fit can be extracted from the
initial decay region, shown in the inset of Fig. \ref{fig8}.
Similar power-law behavior has been found in colloidal gel
simulations \cite{Lodge99}. We are further convinced of this
approach by the apparent change of slope of $G_{eq}'$ at the point
where $G_{eq}$ becomes non-zero. $G_{eq}'$ begins acquiring
non-zero values at higher $T$ (near $T_{MC}$=0.51, the idealized
MCT temperature). In other words, apparently ``solid'' glasses can
have an almost liquid-like response at longer times, provided the
temperature is just above or just below $T_G$. This is perhaps due
to some degree of non-ergodicity in the system. We cannot put a
precise value on the timescale of the long-term rigidity, only
that it is much greater than that of the simulation.

\subsection{The shear modulus from a finite deformation}

We also confirm our $G_{eq}$ results by checking whether or not
they represent the shear modulus $\mu $ in its regular sense,
\textit{i.e.} the response to a shear deformation. $\mu$, the
value of $G_{eq}'$ on our short simulation time scales and the
value of $G_{eq}$ by fit are shown in Fig. \ref{fig9}. This
preliminary comparison is promising, since both $G_{eq}$ and $\mu
(\varepsilon $=0.1) begin acquiring non-zero values at the same
temperature $T_1$ defined in the previous section: they are both
good indicators of the onset of true rigidity. Clearly $\mu
(\varepsilon $=0.1) is much smaller than $G_{eq}$: while $G_{eq}$
calculates the shear modulus via the internal fluctuations, $\mu $
requires external constraints to be imposed, thus altering the
system. As such, we look for the limit $\mu (\varepsilon \to 0$).
Fig. \ref{fig10} shows the function $\mu (\varepsilon )$ for a set
of samples at three given temperatures. This calculation was done
by ensuring that the stress had reached a plateau before
evaluating $\mu $, via Eq. (\ref{mu}). Clearly, this glass
displays highly non-linear behavior and we have, as of yet, no
function with which to fit the points. The limit $\varepsilon \to
0$ is computationally prohibitive to reach. The error in $\mu$ as
seen from Eq. \ref{mu} grows as $1/\epsilon$. Fig. \ref{fig10}
does, however, indicate that $\mu (\epsilon \rightarrow 0) $
likely tends to $G_{eq}$. It also suggests that the regime of
elastic deformation for these glasses is approximately 2 to 3 \%.

\subsection{The short time diffusion coefficients}

\begin{figure}
\resizebox{3.25in}{2.5in}{\includegraphics{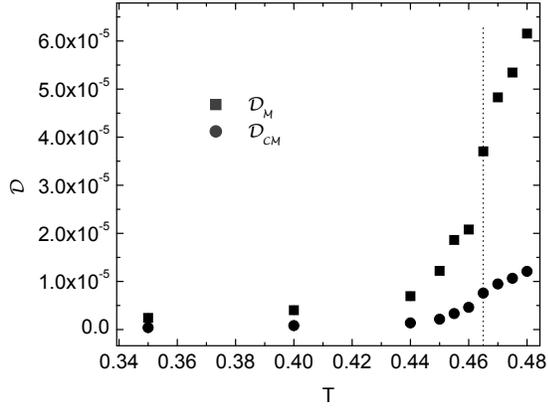}}
\caption{Average short-time diffusion of monomers (${\cal D}_M$)
and centers of mass of chains (${\cal D}_{CM}$) around $T_{G}$.
The substantial increase in ${\cal D}_{M}$ begins slightly below
$T_{G}$ (indicated by the dotted line).}\label{fig5}
\end{figure}

\begin{figure}
\resizebox{3.25in}{2.5in}{\includegraphics{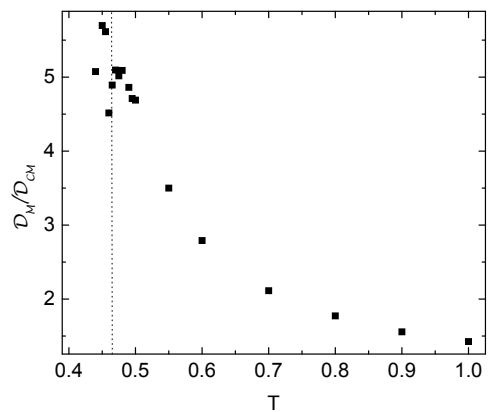}}
\caption{Cage effect as seen by dividing the monomer by the chain
diffusion constants at short times. The maximum occurs near
$T_{G}$ (dotted line), as the cage progressively ``closes
in.''}\label{fig6}
\end{figure}

\begin{figure}
\resizebox{3.25in}{2.5in}{\includegraphics{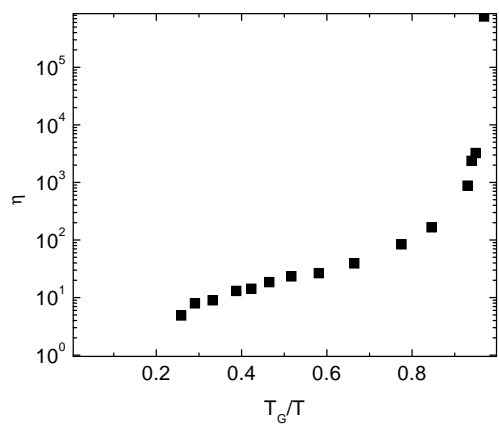}}
\caption{Glassy behavior of the polymer melt shown in a typical
Arrhenius plot, indicating that the glass is primarily
fragile.}\label{fig7}
\end{figure}

\begin{figure}
\resizebox{3.25in}{2.5in}{\includegraphics{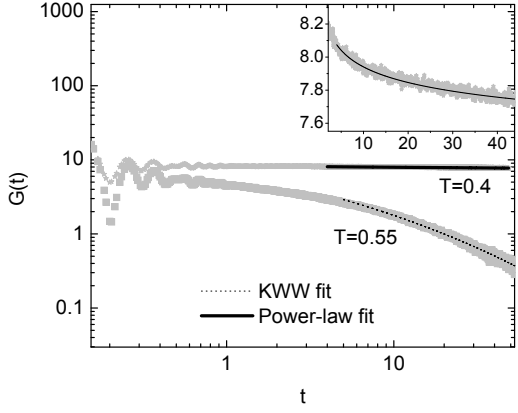}}
\caption{Stress autocorrelation function $G(t)$ decay above and
below $T_{G}$, as calculated from Eq. (\ref{G}), along with the
corresponding KWW and power-law fits from Eqs. (\ref{stretched})
and (\ref{Gpower}). Note the very slow power-law decay. Inset:
Initial decay of the power-law for $T=0.4$ shown on a linear
plot.}\label{fig8}
\end{figure}

\begin{figure}
\resizebox{3.25in}{2.5in}{\includegraphics{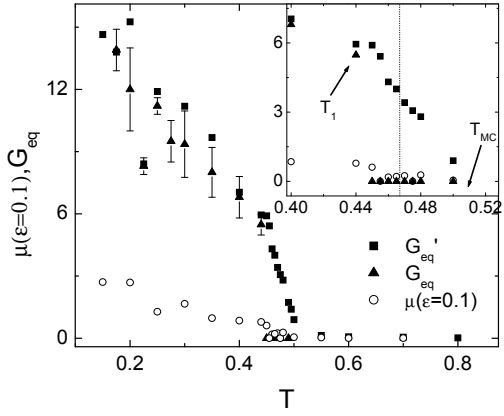}}
\caption{Various approaches to determining the shear modulus.
$G_{eq}'$ is simply the value of $G(t)$ at $t=150$ time units (the
rigidity on very short timescales), while $G_{eq}$ is obtained by
fitting the function to Eq. (\ref{Gpower}) at $t\to\infty$ .
Finally, $\mu$ is calculated from Eq. (\ref{mu}) via a
non-negligible deformation of the sample. The inset shows the
behavior around the GT (dotted line) with the locations of $T_1$
(the onset of long-term rigidity) and $T_{MC}$ (the MCT
temperature) indicated by arrows.}\label{fig9}
\end{figure}

\begin{figure}
\resizebox{3.25in}{2.5in}{\includegraphics{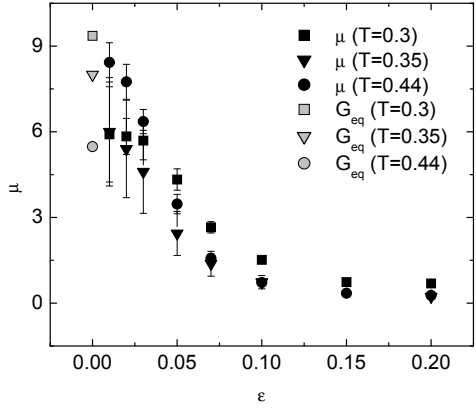}}
\caption{Shear modulus $\mu$ as a function of shear for three
temperatures below the GT. We interpret the $G_{eq}$ as the
zero-shear limit to check whether these two methods agree. Note
how the statistics get much worse at small
$\varepsilon$.}\label{fig10}
\end{figure}

To gain more insight into the structural behavior of the melt near
the glass transition we look at the short-time diffusion
coefficients ${\cal D}_{M}$ and ${\cal D}_{CM}$ of the monomers
and center of mass of chains respectively, defined in Section
\ref{diff}. The measurements reflect, on average, how much local
motion is possible for a given particle. As expected, as we
decrease the temperature,  there is decreased local diffusion as
we approach the RT and GT (see Fig. \ref{fig5}). In addition we
notice that, in the glassy regime, the monomers maintain
significantly more short-time mobility than the chains close to
$T_G$. The chains, on the other hand, are more constrained. This
is further illustrated by the behavior of the ratio ${\cal
D}_{M}/{\cal D}_{CM}$ with temperature. Fig. \ref{fig6} reveals
that monomer motion becomes increasingly dominant as we approach
$T_{G}$ from above. There could be enough free volume for some
monomers to move short distances \cite{Colby00}. This has been
illustrated by the cage effect \cite{Bennemann99}: as the chains
become immobilized near $T_{G}$, the monomers maintain some degree
of mobility within their ``cage'' formed by the neighbors. This
is, in fact, an important precursor to the system becoming not
only ``slow,'' but also rigid. At $T_G$ the monomers still have
sufficient mobility to absorb deformations. This may explain why
the onset of shear rigidity is below $T_G$.

\section{Discussion}

We have determined the glass transition temperature for our system
of freely jointed chains interacting with van der Waals
interactions through the change of slope of the packing fraction
with temperature, or the change in the value of the volume
expansivity. This led to a value of $T_{G}= 0.465 \pm 0.005$. At
the glass transition the heat capacity at constant pressure $C_P$
exhibits a sharp rise with increasing temperature. The above
determined $T_G$ occurs in the middle of the rise. As seen in
Table I a number of other temperatures have been obtained in our
study of the onset of rigidity brought about by the kinetic arrest
occurring near $T_G$. The beginning of the rise in $C_P$ coincides
with $T_{1}= 0.44 \pm 0.01$ the onset of rigidity point as
measured by $G_{eq}$ and $\mu$, and the end of the rise with
$T_{MC}= 0.51 \pm 0.02$, the MCT critical temperature, the
divergence point in the viscosity predicted by mode coupling
theory. These coincidences do not appear fortuitous. The rapid
rise in $C_P$ suggests a temperature interval where the
configuration space explored by the glass in the making changes
rapidly. The end of the rise may very well be the temperature at
which thermally activated processes end and the assumptions of
idealized MCT become valid. $T_{MC}$ is expected to be higher than
$T_G$ \cite{Angell98}. Below $T_{MC}$, the liquid can still flow
with the help of thermally activated hops. The temperature of
rigidity onset was determined by studying the behavior of the
viscosity $\eta$ and the shear modulus upon cooling. The
divergence point of $\eta$ is obtained by extrapolation, while the
onset of rigidity is measured fairly close to the transition. So
we will begin with the latter. At $T_G$ by all indications the
system has no long-time shear resistance. Both $G_{eq}= \lim_{t
\rightarrow \infty} G(t) $ and $\mu$, obtained by an actual
deformation of the system, are still zero. $G'_{eq}=G(t\approx
150)$, a short time shear modulus becomes non-zero around
$T_{MC}$. But $G_{eq}$, its long time limit only sets in at
$T_{1}= 0.44 \pm 0.01$. At the same temperature $\mu$ becomes
non-zero, although the values of $\mu$ are not the same as those
of $G_{eq}$ for typical deformations of $\epsilon = 0.05$ or $0.1$
(see Figs. \ref{fig8}and \ref{fig9}). We will elaborate on these
differences later in the discussion. The important point is that
they agree on the temperature of rigidity onset. On the other
hand, the divergence point of the viscosity $\eta$ was determined
using several models, two of which incorporate thermally activated
processes, which become operative below $T_{MC}$. They predict
divergences at $T_{0}= 0.41 \pm 0.02$ from a fit to the VFT law
(see Eq. (\ref{etaVFT})) and $T_C = 0.422 \pm 0.006$ from the
Colby form (see Eq. (\ref{Colby}). One should, however, note that
these are obtained by extrapolating from the lowest temperature at
which $\eta$ have been measured, $0.49$. The true states at $T_0$
or $T_C$ are not attained due to the dramatic slowing-down below
$T_G$. The extrapolated values are smaller than $T_1$, but not
inconsistent with it. They are not expected to be as reliable
because they are not measured directly. The fact that $T_1$ is
smaller than $T_G$ determined from the change in volume
expansivity raises the question about the true location of the
glass transition. Does this mean that the beginning of the rise in
$C_P$ should be used as the true onset of the glass transition, or
is it, that at $T_G$, although there is structural arrest,
sufficient free volume remains to allow for small deformations at
no cost of energy?

A number of aspects of this study are worth commenting on further.
First there is always a time scale associated with rigidity, as it
is well known that the shear response depends on frequency. What
we are trying to determine is the static shear modulus, or zero
frequency limit. The long time associated with this modulus, as
measured, is still short compared with temperatures related with
aging. We started measuring quantities when the system showed no
sign of evolving. In other words aging is on a different time
scale than stress relaxation times. An unrelated issue on time
scale is the definition itself of $T_G$ which is linked to the
experimental time scale. The glass makers, as mentioned in Section
\ref{glasstransition}, use a value of $10^{13}$ Pa$\cdot$s for the
viscosity at $T_G$. This corresponds to the experimental time
scale of minutes (or hours) \cite{Rao} while a computer
experiment, under the best of conditions, occurs over nanoseconds,
so about $10^{11}$ smaller. It is intriguing to note that the VFT
fit and the Colby fit yield viscosities at $T_G$ of the order of
$10^7$ in our units, which with $G_{\infty}$ of the order of
$10^2$ gives a relaxation time of the order of $10^5$ at $T_G$,
longer than the time scale of our computer experiments, closer to
$10^4$, or $10^6$ time steps.

Another point is the nature of the onset of rigidity. What happens
at $T_1$? Studies underway \cite{Wallace} show that the
distribution of displacements on all time scales studied is
unimodal as previously observed \cite{Vollmayr02}, so there is no
separate long-lived rigid backbone that accounts for the
resistance to applied shear. Second, the local structure of the
system remains identical, as expected from a GT. In other words,
there is no evidence of any clusters of tightly-packed particles.
The average distance to the nearest-neighbors of a given particle
does not change appreciably with temperature. The concept of
dynamical heterogeneity may be more applicable to this system
\cite{Gebremichael01,Vollmayr02}, in that the lack of mobile
clusters on fairly long timescales below $T_1$ accounts for
rigidity. Instead of having covalent bonding, the van der Waals
rigidity in our system may arise from ``jamming'' constraints
produced by mechanisms such as the cage effect, which relies
primarily on the presence of stiff chains to prevent motion.

The fact that the appearance of long-term rigidity at $T_1=0.44$
coincides with the bottom of the characteristic ``dip'' in the
heat capacity curve (Fig. \ref{fig3}), is indicative of an
increase in system stability as the temperature is lowered. In
network glasses, the number of floppy modes in the system
determines the jump in $C_P$ (also related to the fragility of the
glass) \cite{Naumis00}. And these floppy modes start appearing at
$T_1$. Other approaches have been able to associate $C_P$ with the
change in entropy of constraint breaking \cite{AngellBook}.
Although these approaches rely on physical bonds as constraints,
the same principles could apply to our system. Furthermore, we
notice from Fig. \ref{fig4} that ${\cal D}_{CM}$ is very small at
$T_1$. Due to the noise in the data, it is not likely that we
could resolve the entropic component to the rigidity if it exists.
In other words this rigidity transition is probably mechanical in
nature.

\begin{figure}
\resizebox{3.25in}{2.5in}{\includegraphics{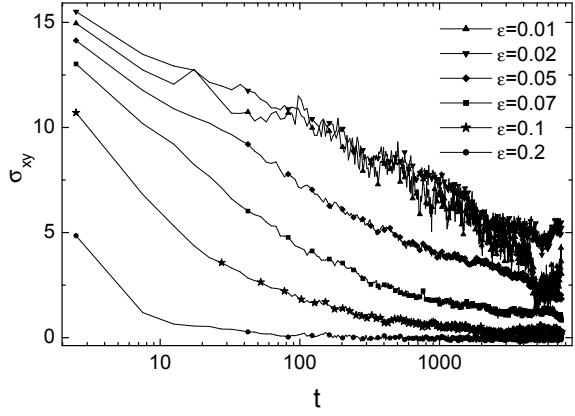}}
\caption{Decay of of the shear stress $\sigma _{xy}$ after various
initial shear deformations at $T=0.3$. We can see both the initial
stress and the shape of the subsequent decay in the system. Note
that for large $\varepsilon$ there is very little initial stress
imparted, followed by a fast decay.}\label{fig11}
\end{figure}

\begin{figure}
\resizebox{3.25in}{2.5in}{\includegraphics{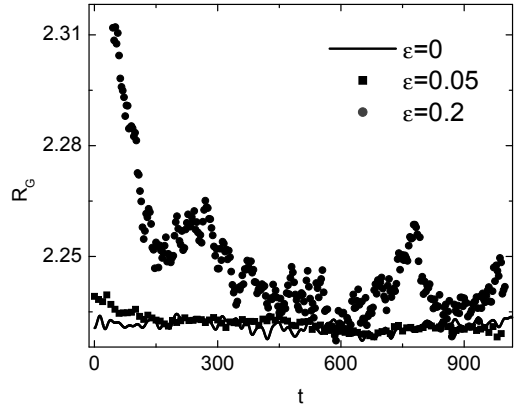}}
\caption{Evolution of the average radius of gyration, $R_{G}$, of
the polymer chains, after a large  ($\epsilon =0.2$) and small
($\epsilon =0.05$) deformation at $t=0$ together with the
undeformed, ``equilibrium'' $R_{G}$  ($\epsilon
=0$).}\label{fig14}
\end{figure}

Finally we would like to comment on the difference between
$G_{eq}$ and $\mu$, the latter obtained from the application of an
instantaneous simple shear to the system. The non-linear behavior
of Fig. \ref{fig10} is further explained in Fig. \ref{fig11}. Both
the initial stress and the degree of relaxation are highly
dependant on $\varepsilon$ (Fig. \ref{fig11}). For larger
shearing, the initial stress is smaller and the subsequent decay
is very fast. A collapse was realized by plotting the stress
scaled to its initial value versus time scaled with a relaxation
time equal to the time to reach half the stress.  The relaxation
times obtained follow the same trend as µ. The shearing is likely
irreversible at large strain. In effect, large deformations appear
to have a substantial effect on the structure of the system,
perhaps changing the shape of the energy landscape and causing the
system to be in a different local minimum. Small shear, however,
does not generally have the same effect and a larger residual
stress remains as the system tries to return to its original
configuration (the same energy well). Having described the basic
behavior associated with shearing, we can alternatively examine
the local ``polymeric'' contribution to the mechanical relaxation
of the glass. Considering Fig. \ref{fig14}, we can see that for
small $\varepsilon$, the chains are slightly stretched, but can
easily recover their equilibrium $R_G$ value, where $R_G$ is the
average radius of gyration of the chains. For larger strains, the
initial deformation of chains is disproportionately high and there
can be no recovery on intermediate time scales. We can interpret
the degree of recovery in the chain as the main factor
characterizing the viscoelastic behavior in polymer melts (often
represented by spring-dashpot models), since chain length has a
direct, substantial effect on relaxation mechanisms. Furthermore,
rigidity should occur earlier in a system of longer chains, since
each particle will have, on average, fewer degrees of freedom.

\section{Conclusion}

Our isothermal compression method has allowed us to better examine
the state of the melt above and below $T_G$. Our studies of
rigidity bring additional insight into the nature of the GT.
Previous studies on structural issues related to the glass
transition have focussed on the divergence of the viscosity. We
provide an additional perspective by looking also at the emergence
of the shear resistance. Whereas the point of divergence of the
viscosity is difficult to assess directly because of structural
arrest, it appears easier to approach the point of onset of the
shear modulus. That both $G_{eq}$ and $\mu$ have the same onset is
encouraging. The work demonstrates the effectiveness of the stress
correlators, from which $G_{eq}$ is obtained, compared to the
application of an external deformation which yields $\mu$. The
$T_{G}=0.465$ determined from the change in packing fraction lies
in the middle of the rapid rise of the heat capacity. With
decreasing temperature, there is a short transition period of
short-term rigidity beginning near $T_{MC}=0.51$ (the end of the
rise in $C_P$), presumably due to a lack of ergodicity, followed
by the appearance of a long-term non-zero shear modulus at
$T_1=0.44$ (the beginning of the rise). I believe that we have
only begun to explore what shear resistance can teach us about the
GT: from non-linear mechanical properties, frequency dependences,
to issues related to aging.

%\newpage

\end{document}